\documentclass[sn-aps]{sn-jnl}

\usepackage{graphicx}%
\usepackage{multirow}%
\usepackage{calligra}
\usepackage{amsmath,amssymb,amsfonts}%
\usepackage{amsthm}%
\usepackage{mathrsfs}%
\usepackage[title]{appendix}%
\usepackage{xcolor}%
\usepackage{textcomp}%
\usepackage{manyfoot}%
\usepackage{booktabs}%
\usepackage{algorithm}%
\usepackage{algorithmicx}%
\usepackage{algpseudocode}%
\usepackage{listings}%
\usepackage{multirow}
\usepackage{subcaption}
\usepackage{placeins}
\usepackage{setspace}
\usepackage{orcidlink}
\usepackage{physics}
\usepackage{svg}
\usepackage{caption,subcaption}
\usepackage{svrsymbols} 
\usepackage{siunitx}
\usepackage{soul}
\DeclareSIUnit\angstrom{\text{Å}}

\usepackage[color=orange!60,textsize=footnotesize]{todonotes}
\DeclareMathAlphabet{\mathcalligra}{T1}{calligra}{m}{n}
\DeclareFontShape{T1}{calligra}{m}{n}{<->s*[2.2]callig15}{}
\newcommand{\scriptr}{\mathcalligra{r}\,}

\begin{document}

\title{Effects of substrate corrugation during helium adsorption on graphene in the grand canonical ensemble}


\author[1,2]{\fnm{Gage} \sur{Erwin} \orcidlink{0000-0002-2219-5756}}
\author[1,2,3]{\fnm{Adrian} \sur{Del Maestro} \orcidlink{0000-0001-9483-8258}}
\affil[1]{\orgdiv{Department of Physics and Astronomy}, \orgname{University of Tennessee, Knoxville}, \orgaddress{\state{Tennessee}, \postcode{37996}, \country{USA}}}

\affil[2]{\orgdiv{Institute for Advanced Materials and Manufacturing}, \orgname{University of Tennessee, Knoxville}, \orgaddress{\state{Tennessee}, \postcode{37996}, \country{USA}}}

\affil[3]{\orgdiv{Min H. Kao Department of Electrical Engineering and Computer Science}, \orgname{University of Tennessee, Knoxville}, \orgaddress{\state{Tennessee}, \postcode{37996}, \country{USA}}}


\abstract{Adsorption of ${}^4$He on graphene substrates has been a topic of great interest due to the intriguing effects of graphene corrugation on the manifestation of commensurate solid and exotic phases in low-dimensional systems. In this study, we employ worm algorithm quantum Monte Carlo to study helium adsorbed on a graphene substrate to explore corrugation effects in the grand canonical ensemble. We utilized a Szalewicz potential for helium-helium interactions and a summation of isotropic interactions between helium and carbon atoms to construct a helium-graphene potential. We implement different levels of approximation to achieve a smooth potential, three partially corrugated potentials, and a fully ab initio potential to test the effects of corrugation on the first and second layers. We demonstrate that the omission of corrugation within the helium-graphene potential could lead to finite-size effects in both the first and second layers. Thus, a fully corrugated potential should be used when simulating helium in this low-dimensional regime.}




\maketitle

\section{Introduction}\label{sec1}

The adsorption of ${}^4$He atoms on a solid substrate has been an intriguing many-body problem for decades, holding the possibility of dimensional crossover as the film thickens, and exhibiting a rich phase diagram \cite{bib1,bib2,bib3,bib4,bib5,bib6,bib7,bib8,bib9}. Previous experimental \cite{bib2,bib10,bib11} and theoretical studies \cite{bib10,bib12,bib13,bib14,bib15,Boninsegni:2020sh} demonstrated that ${}^4He$ on graphite is an ideal substrate for investigating both solid commensurate and incommensurate layers close to the surface, as well as low-dimensional superfluidity in the 2nd and higher adsorbed layers. Additionally, there is a history of intriguing observations reporting the possible coexistence of superfluidity and density wave order \cite{bib18} including experimental studies \cite{Nyki:2017py,bib19,bib8}.  While there is still debate on whether or not this behavior can be realized in simulations \cite{bib15,bib21,bib23, Boninsegni:2020sh}, there is a resurgent interest in the properties of low dimensional adsorbed phases of helium. Adding to this is the discovery of graphene \cite{RevModPhys.81.109}, an atomically thin substrate with reduced (and possibly tunable \cite{bib16}) van der Waals interactions (compared to graphite), which should enhance delocalization and thus quantum effects near the substrate \cite{bib9,bib17,bib23,Kim:2023ss}. Yet, there is still a question of whether the current effective and empirical models used in simulations to describe both helium-graphite and helium-graphene \cite{bib22, Bruch:2010wo, Burganova:2016ht, Yu:2021gr}, constructed from the superposition of ${}^4$He-C interactions, are sufficient to adequately represent the physics of adsorption under experimental conditions, and in particular, what role corrugation plays in the first and higher adsorbed layers.

This problem was previously addressed for graphite by Pierce and Manousakis \cite{bib20}, who studied the effects of corrugation on the first layer of ${}^4$He, concluding that it is required to realize the $\sqrt{3} \times \sqrt{3}$ commensurate solid phase that is well known in experiments, and fully confirmed with quantum Monte Carlo \cite{bib15}.  However, it was less clear whether or not the second adsorbed layer was far enough away from the substrate that corrugation would be relevant, and thus the second layer could be simulated assuming graphene to be featureless. Boninsegni and Moroni recently addressed this question via numerical simulations \cite{bib21} concluding that the effects of surface corrugation on the second layer of ${}^4$He adsorbed on graphite were negligible.

These previous works were all performed in the canonical ensemble, studying the properties of the first and higher adsorbed layers as a function of helium coverage.  In this study, we investigate the effects of corrugation for helium adsorbed on graphene in the \emph{grand canonical ensemble}.  We simulate the adsorption process for helium on graphene as a function of chemical potential, directly related to the experimentally tunable pressure.  This is accomplished via worm algorithm path integral quantum Monte Carlo \cite{bib30,bib31}, which directly operates in the grand canonical ensemble.  We probe corrugation effects analogously to Ref.~\cite{bib21} by systematically increasing the number of symmetry related reciprocal lattice vectors included in a 6-12 Carlos-Cole potential \cite{bib22} taking into account the superposition of isotropic helium-carbon interactions.  

Our results are in agreement with previous studies demonstrating that the existence of a first order transition out of the vacuum to a $\sqrt{3} \times \sqrt{3}$ commensurate solid phase in the first layer \cite{bib17,bib23,bib24,bib25} is reliant on the use of a corrugated potential.  For a smooth potential, we instead observe that the adsorption process extends over a region of chemical potentials with the adsorbed layer remaining compressible. As the chemical potential is increased, and second layer formation becomes energetically favorable above $\sim 0.6$ filling, we find that potential corrugation leads to a slightly more bound first layer which may inhibit fluctuations in the particle number and subsequently reduces the compressibility upon the second layer onset.  This could have implications for the phases and phase transitions of the second layer.

In the remainder of this paper, we first present the microscopic model we employ to characterize the adsorption of helium atoms onto a graphene substrate. We show how varying levels of corrugation are achieved at the level of an empirical helium-graphene potential and briefly describe our quantum Monte Carlo methodology and system parameters.  We present numerical results on the adsorption process for different levels of potential corrugation and directly compare to past theoretical and experimental studies. We analyze finite size effects by considering three different system sizes and consider the extrapolation of the energy per particle to the thermodynamic limit. We discuss implications for the second adsorbed layer and outline some promising future directions of inquiry. All code, scripts and data needed to reproduce the results in this paper are available online \cite{papersCodeRepo} and \cite{del_maestro_2023_10137837}.

\FloatBarrier

\section{Model\label{sec5}}

\begin{figure}[t]
    \centering
    \includegraphics[width=.8\textwidth]{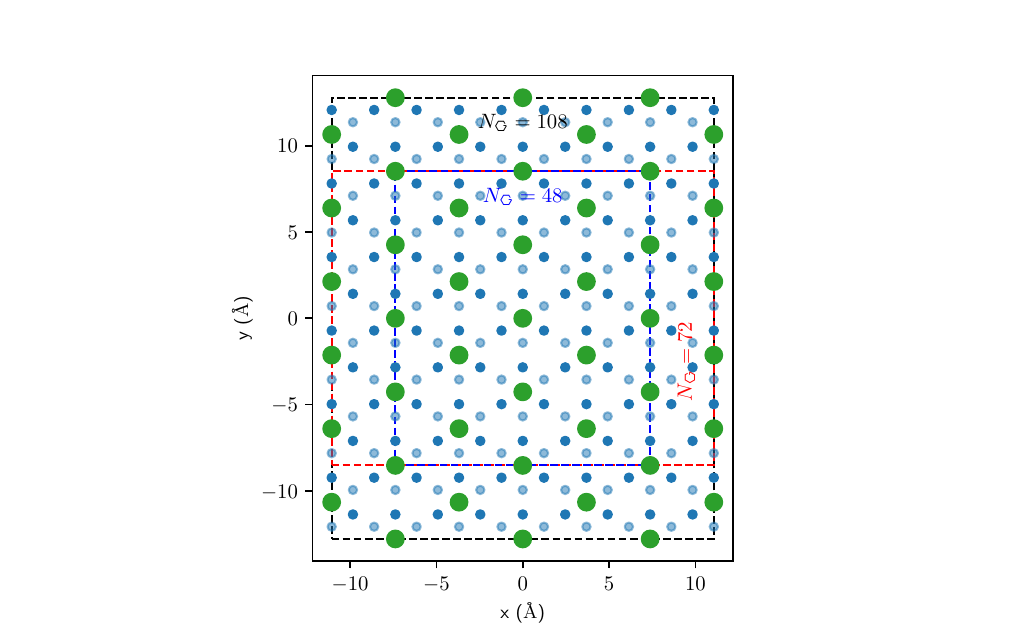}
    \caption{Three system sizes with periodic boundary conditions in the $xy$-plane. Carbon atoms are shown by blue solid circles, and the larger green solid circles indicate the strong adsorption sites corresponding to the C1/3 phase. $N_\graphene$ refers to the number of adsorption sites. Reproduced using graphenetools-py \cite{Graphenetools}.}
    \label{Fig:1}
\end{figure}

We consider $N$ indistinguishable ${}^4$He atoms of mass $m$ proximate to a graphene substrate oriented in the $xy$-plane and fixed at $z=0$. The system is enclosed in a simulation cell of size $L_x \times L_y \times L_z$ with periodic boundary conditions in the $x$ and $y$ directions. Atoms are restricted to only one side of the graphene sheet corresponding to $z>0$ through a hard wall located at $z=L_z$ defined by:
\begin{equation}
    \mathcal{V}_{\rm wall}(z) = \frac{\mathcal{V}_{\rm He -\graphene}(\vb*{r}_{\rm min})}{1 + 
    \mathrm{e}^{(L_z-r_{\rm vdW}-z))/\Delta}}.
\label{eq:Vwall}
\end{equation}
The parameters were chosen to approximate the hard-core part of helium-graphene
potential and we take $\vb*{r}_{\rm min} = a_0(\sqrt{3}/2,1/2,1)$ (here $a_0 =
\SI{1.4}{\angstrom}$ is the carbon-carbon distance) leading to
$\mathcal{V}_{\rm He -\graphene}(\vb*{r}_{\rm min}) / k_{rm B} \sim
\mathrm{O}(10^5)~\si{\kelvin}$ (where $k_{\rm B}$ is the Boltzmann constant).  Other parameters include $r_{\rm vdW} \approx \SI{1.4}{\angstrom}$, the van der Waals radius of helium, and $\Delta = \SI{0.05}{\angstrom}$, which defines the rapidness of the wall onset.  In a previous study \cite{Yu:2021gr} we confirmed that quantitative modifications of the potential do not affect the physics of adsorption provided $L_z \gtrsim \SI{6}{\angstrom}$. In this study, we fix $L_z = \SI{10}{\angstrom}$ and select three transverse cell-size commensurate with periodic boundary conditions for graphene:$L_x = \SI{14.757}{\angstrom}$ and $L_y = \SI{17.04}{\angstrom}$ admitting 48 strong adsorption sites, $L_x = \SI{22.136}{\angstrom}$ and $L_y = \SI{17.04}{\angstrom}$ admitting 72 strong adsorption sites, and $L_x = \SI{22.136}{\angstrom}$ and $L_y = \SI{25.56}{\angstrom}$ admitting 108 strong adsorption sites as shown in Fig.~\ref{Fig:1}.

The adsorption process is governed by the $N$-body Hamiltonian for a system of ${}^4$He atoms with mass $m$ interacting with a graphene substrate: 
\begin{equation}\label{Equation:1}
    \mathcal{H} = -\frac{\hbar^2}{2m}\sum_{i=1}^{N}\nabla_i^2 + \sum_{i=1}^{N}\mathcal{V}_{\rm He-\graphene}(\mathbf{r}_i) + \sum_{i<j}\mathcal{V}_{\rm He-He}(\mathbf{r}_i-\mathbf{r}_j)
\end{equation}
where we neglect 3-body interactions (expected to be weak for single layer materials \cite{McLachlan1964}) and the $i^{th}$ atom is located at $\mathbf{r}_i = (x_i,y_i,z_i)$. The interaction term $\mathcal{V}_{He-He}$ is taken to be of the Szalewicz form \cite{bib27, bib28} and is known to high accuracy. The second term includes all interactions between the helium atoms and the graphene substrate. Here, we employ the same potential as in Ref.~\cite{bib29} derived from the sum of isotropic 6-12 Lennard-Jones potentials with parameters updated via graphene polarization calculations \cite{bib16}.  The use of an isotropic (as opposed to anisotropic) helium-carbon dispersion interaction is employed due to the lack of experimental scattering data that could be used to fit the additionally needed parameters for graphene.  The resulting corrugated adsorption potential is given by:
\begin{equation}\label{Equation:2}
    \begin{split}
    \mathcal{V}_{\rm He-\graphene}(\vb{r}_i) = &\varepsilon\sigma^2\frac{4\pi}{A}\biggl\{\biggl[ \frac{2}{5}\biggl(\frac{\sigma}{z}\biggr)^{10}-\biggl(\frac{\sigma}{z}\biggr)^{4}\biggr]\\ &+ \sum_{\mathbf{g}\ne 0}\sum_{k=1}^q \mathrm{e}^{\mathrm{i}\mathbf{g}\cdot[\mathbf{m}_k + \scriptr_i]}\biggl( \frac{1}{60}\biggl(\frac{g\sigma^2}{2z}\biggr)^{5}K_5(gz)-\biggl(\frac{g\sigma^2}{2z}\biggr)^{2}K_2(gz)\biggr)\biggr\}.\\
     \end{split}
\end{equation}
In this expression, $A= \SI{5.239}{\angstrom^2}$ is the area of the unit cell, $\scriptr_i = (x_i,y_i)$ are the coordinates of the ${}^4$He atom in the xy-plane, and the sum runs over all of the graphene reciprocal lattice vectors, $\mathbf{g} = l_1\mathbf{G_1} + l_2\mathbf{G_2}$, where 
\begin{equation}
    \vb{G}_1 = \frac{2\pi}{3a_0}\qty(\sqrt{3},1), \qquad 
    \vb{G}_2 = \frac{2\pi}{3a_0}\qty(-\sqrt{3},1).
\label{eq:grapheneG}
\end{equation}
The real-space basis vectors are:
\begin{equation}
    \vb{m}_1 = \frac{a_0}{2}\qty(\sqrt{3},1), \qquad 
    \vb{m}_2 = a_0\qty(0,1).
\label{eq:grapheneBasis}
\end{equation}
and $K_n(\cdot)$ are modified Bessel functions with asymptotic behavior
$\exp(-t)$ for $t \gg 1$.  We employ hybridized helium-carbon Lennard-Jones
parameters $\epsilon = \SI{16.968}{\kelvin}$ and $\sigma =
\SI{2.641}{\angstrom}$ determined for graphene \cite{bib16}, which differ
slightly from their counterparts for graphite \cite{bib15}.  The first term in Eq.~(\ref{Equation:2}) corresponds to the far-field smooth potential ($\vb{g} =0$) which only depends on the distance form the membrane and is equivalent to what a ${}^4$He atoms would experience from a uniform 2D slab with the same density as graphene.  The effects of corrugation can be explored by systematically including additional sets or ``shells'' of symmetry-related reciprocal lattice vectors in the second term of $\mathcal{V}_{\rm He-\graphene}$ all having the same magnitude $\abs{\vb{g}}$. As this term converges rapidly with increasing $\abs{\vb{g}}$, we need only consider 5 distinct potentials corresponding to the 4 shells defined in Table~\ref{Table:1} and the fully corrugated potential, where the sum is only restricted once we reach double numerical precision.  
\begin{table}[h]
    \renewcommand{\arraystretch}{1.5}
    \setlength\tabcolsep{12pt}
    \begin{tabular}{@{}lll@{}} 
        \toprule
        \textbf{n$^{th}$ shell} & \textbf{Integer multiples $(l_1, l_2)$ for reciprocal lattice vectors $\mathbf{g} = l_1\mathbf{G_1} + l_2\mathbf{G_2}$} \\
        \midrule
    0 & (0,0) \\
    1 & (0,1), (1,0), (1,1), (0,-1), (-1,0), (-1,-1) \\
    2 & (1,2), (2,1), (-1,1), (-1,-2), (-2,1), (1,-1) \\
    3 & (2,2), (2,0), (0,2), (-2,-2), (-2,0), (0,-2) \\
        \bottomrule
    \end{tabular}
    \caption{Reciprocal Lattice Vectors corresponding to different levels of corrugation for the helium-graphene potential in Eq.~(\ref{Equation:2}). \label{Table:1}}
    \label{reciprocaltable}
\end{table} 

\begin{figure}[t]
    \centering
    \includegraphics[width=1\textwidth]{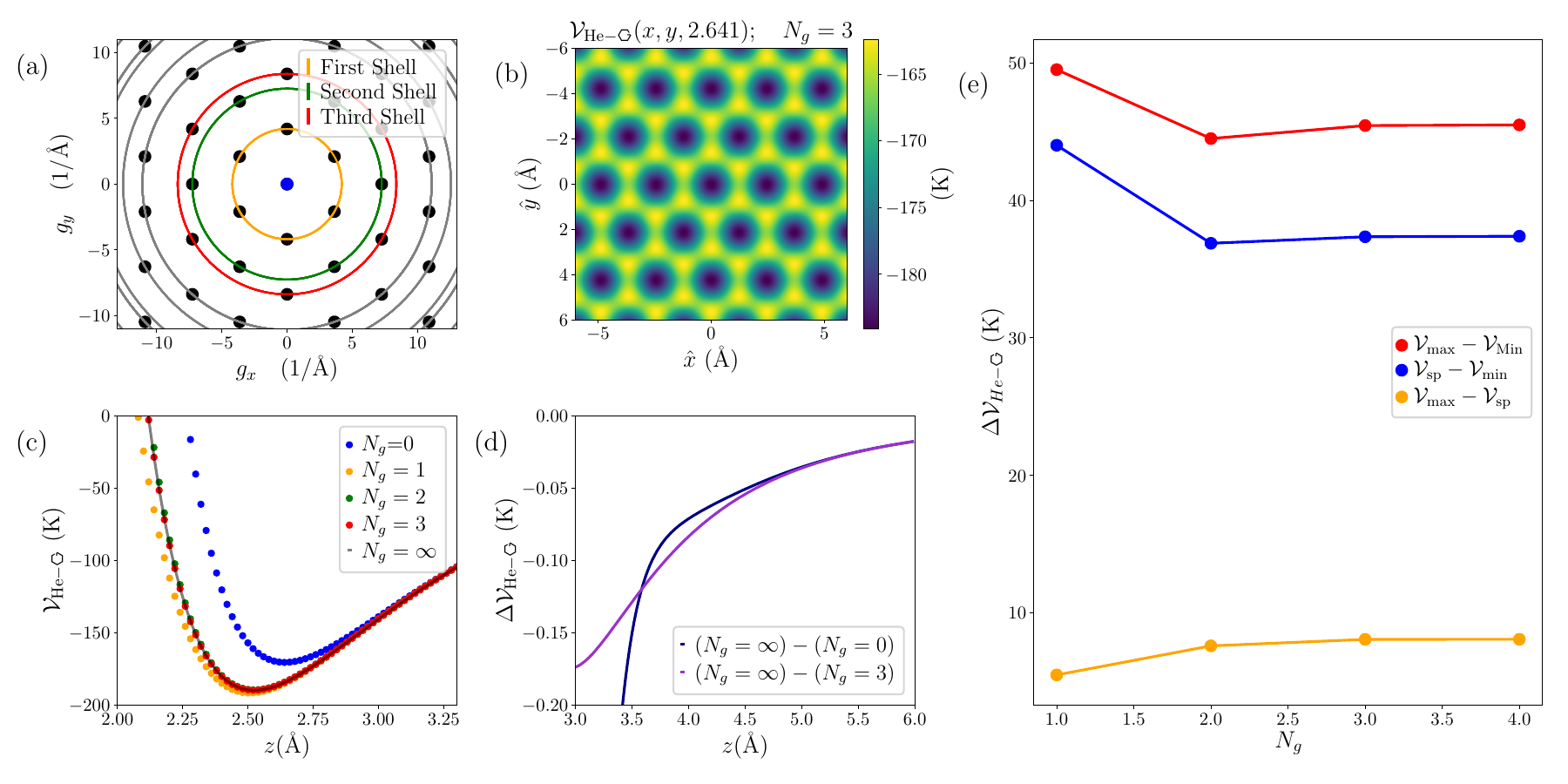}
    \caption{(a) The first four reciprocal lattice vector shells (referred to as $N_g$) are depicted in varying colors. (b) The adsorption potential $\mathcal{V}_{\rm He-\graphene}$ in the $xy$-plane, at fixed distance of $z=\SI{2.641}{\angstrom}$. (c) The effects of corrugation on a ${}^4$He atom located at $\vb{r}=(0,0,z)$ near the graphene membrane. Shells up to $N_g=\infty$ are included, with the latter indicating accuracy of order $10^{-16}~\text{K}$.  Deviations between this result and a smooth ($N_g=0$) and partially corrugated ($N_g=3$) potential are quantified in panel (d). (e) The effects of corrugation on the energy gap between the maxima and minima, the saddlepoint and minima, and maxima and saddlepoint which can be seen in spatial dependence of the adsorption potential panel (b).
    }
    \label{Fig:2}
\end{figure}

One aim of this work is to determine the minimum number of reciprocal lattice vectors required to accurately represent the adsorption potential and some details of the potential are included in Fig.~\ref{Fig:2}.
In panel (a) the different reciprocal lattice shells are identified, while panel (b) shows the corrugated adsorption potential above the graphene sheet at a height of $z=\SI{2.641}{\angstrom}$ corresponding to the distance of the potential minimum directly above a strong adsorption site (center of the graphene hexagon).  The $z$-dependence for different levels of corrugation is seen in panel (c) showing that the smooth potential will lead to subtantially different adsorption behavior at small distances.  This effect is quantified in panel (d) showing deviations for $z \le \SI{3.5}{\angstrom}$. The main features of the corrugated adsorption potential (not present for $\vb{g}=0$) that affect physical phenomena near the membrane include local minima, maxima and the saddlepoints between them.  
The minimum ($\mathcal{V}_{\rm min}$) occurs at the center of the hexagon formed by carbon atoms, the maximum ($\mathcal{V}_{\rm max}$) directly above a carbon atom, and the saddle point ($\mathcal{V}_{\rm sp}$) in between carbon atoms. These features are reproduced in the probability density (wavefunction) of adsorbed helium atoms (see e.g. Ref.\cite{bib20,bib23,Yu:2021gr}). The effects of corrugation on these features is shown in Fig.~\ref{Fig:2}e where a non-monotonic convergence is observed as the shell is increased, a result of the modified Bessel function in Eq.~\ref{Equation:2}.  

\FloatBarrier
\section{Methodology}\label{sec4}

As mentioned in the introduction, we employ worm algorithm path integral quantum Monte Carlo to simulate finite temperature adsorption phenomena defined by Eq.~\ref{Equation:1} in the grand canonical ensemble \cite{bib30,Herdman:2014kp,bib31}.  The path integral framework allows for the stochastically exact computation of physical observables via:
\begin{equation}
    \langle \hat{\mathcal{O}} \rangle = \frac{1}{\mathcal{Z}}
    \Tr\qty[\hat{\mathcal{O}}e^{-\beta(\hat{\mathcal{H}}-\mu N)}]
\end{equation}
where $\mathcal{Z} = \Tr \mathrm{e}^{-\beta(\hat{\mathcal{H}}-\mu \hat{N})}$ is grand partition
function for chemical potential $\mu$ and $\beta = 1/T$ (we work in units where
$k_{\rm B} = 1$). We are interested in the effects of corrugation on the
adsorption process quantified by the average filling fraction $n \equiv 
\expval{\hat{N}}/N_{\graphene}$ where $N_{\graphene}=48$ is the number of strong
adsorption sites, the energy per particle $E/N \equiv
\expval{\hat{\mathcal{H}}/N}$, and the compressibility, $\kappa = \dv{\rho}{\mu} = \expval{(N-\expval{N})^2}/{(T V)}$, where $V = L_x L_y L_z$ and $\rho = \expval{N}/V$. Structural information on the thickness and properties of the adsorbed helium film can be determined from the linear density:
\begin{equation}
\rho(z) = \expval{\sum_{i=1}^N \delta(z_i - z)} \propto \iint \dd{x}\dd{y}\abs{\Psi(x,y,z)}^2
\end{equation}
where any corrugation effects have been averaged over the sheet. Thus, by performing simulations at different chemical potentials (a proxy for the experimentally tunable pressure) at fixed temperature, we can explore adsorption phenomena as a function of the number of reciprocal lattice shells ($N_g$) included in the corrugation potential.  This includes corrugation effects on a first order transition from vacuum to a single adsorbed layer, as well as promotion to a second layer as the chemical potential is increased. 

Simulations were performed using an open source path integral Monte Carlo code maintained by the authors \cite{bib32} for different $N_g$ using the parameters included in Table~\ref{Table:2}. To investigate the interplay of potential corrugation and finite size effects we employed periodic cell dimensions in the $x$ and $y$ directions that accommodated $N_\graphene = 42$, $72$, and $108$ strong adsorption sites. 
\begin{table}[h]
    \renewcommand{\arraystretch}{1.5}
    \setlength\tabcolsep{12pt}
    \begin{tabular}{@{}lll@{}} 
        \toprule
        \textbf{Parameter} & \textbf{Value} \\
        \midrule
Temperature &  $T = \SI{1}{K}$ \\
Chemical Potential & $\mu =$~\SIrange{-125}{-15}{\kelvin}; $\Delta \mu = \SI{1}{\kelvin}$ \\
Imaginary time step & $\tau = \SI{0.003130}{\kelvin^{-1}}$ \\
Cell Size $(N_\graphene = 48)$  & $L_x \times L_y \times L_z = \SI{14.757}{\angstrom} \times \SI{17.04}{\angstrom} \times \SI{10.0}{\angstrom}$ \\
Cell Size $(N_\graphene = 72)$ & $L_x \times L_y \times L_z = \SI{22.136}{\angstrom} \times \SI{17.04}{\angstrom} \times \SI{10.0}{\angstrom}$ \\
Cell Size $(N_\graphene = 108)$ & $L_x \times L_y \times L_z = \SI{22.136}{\angstrom} \times \SI{25.56}{\angstrom} \times \SI{10.0}{\angstrom}$ \\
        \bottomrule
    \end{tabular}
\caption{Simulation Parameters}\label{Table:2}
\end{table}

\FloatBarrier

\section{Results}\label{sec6}

\subsection{Corrugation Effects in the First Adsorbed Layer}

\begin{figure}[t]
    \centering
        \centering
        \includegraphics[width=0.49\textwidth]{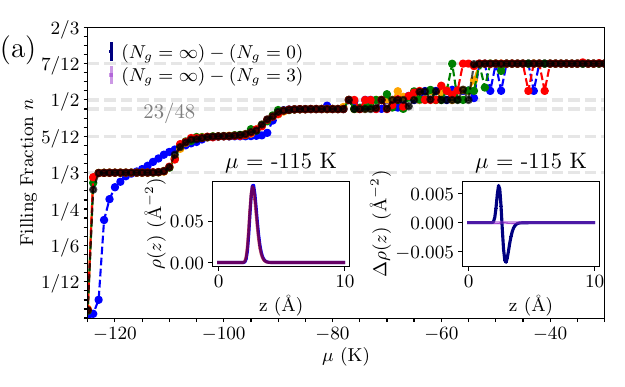}
        \includegraphics[width=0.48\textwidth]{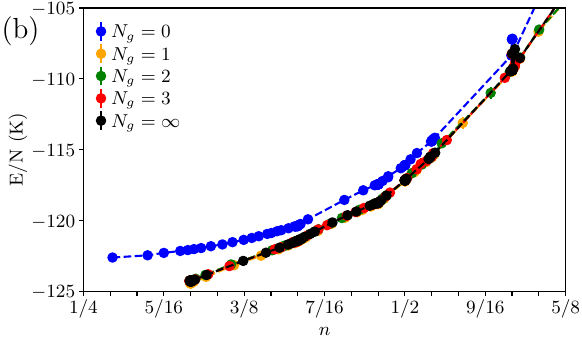}
    \caption{(a) Quantum Monte Carlo results for the filling fraction $n = \expval{N}/N_{\graphene}$ as a function of chemical potential $\mu$ (points) for different levels of potential corrugation as defined by the number of reciprocal lattice shells $N_g$ included in the adsorption potential in Eq.~\ref{Equation:2}. Both panels use the same legend and lines are guides to the eye.  The left inset shows the linear density perpendicular to the graphene membrane indicating the existence of a single well-defined layer at $\mu = \SI{-115}{\kelvin}$. The right inset quantifies the structural difference when corrugation is included, with the legend corresponding to the different values of  $N_g$. (b) The energy per particle $E/N$ vs. filling fraction $n$ in the grand canonical ensemble for different levels of corrugation $N_g$.
    }
    \label{Fig:3&4}
\end{figure}

We begin our discussion of simulation results by focusing on the adsorption of the first layer of helium on graphene. It is known from previous simulations \cite{bib23,Yu:2021gr}, that the first layer adsorbed out of the vacuum is a $\sqrt{3} \times \sqrt{3}$ commensurate solid phase stabilized by the presence of corrugation. This incompressible phase consists of helium  
atoms occupying 1/3 of the strong binding sites on a triangular lattice (hexagon centers) with constant $\sqrt{3}a_0$ and axes rotated by \SI{30}{\degree} with respect to the original graphene triangular lattice. This first order layering transition to 1/3 filling can be clearly seen in Fig.~\ref{Fig:3&4}(a) for all $N_g > 0$ with onset occurring near $\mu = \SI{-126}{\kelvin}$ and $\mu = \SI{-128}{\kelvin}$ for the smooth and corrugated potential respectively.

The large region of stability (as a function of chemical potential) is further confirmed by the energy per particle in panel (b) where a minimum clearly occurs at $n = 1/3$ corresponding to $\expval*{\hat{N}} = 16$ for this particular cell.  By the time we have $N_g=1$, corrugation effects appear to be fully saturated, while there may be some residual rounding and enhanced finite size effects at the first layer adsorption transition.  The situation is drastically different for the smooth potential ($N_g=0$) where no sharp transition is seen, and instead a broad onset region of adsorption appears above $\mu=\SI{-125}{\kelvin}$ and extends to nearly $\SI{-100}{\kelvin}$.  The energy is consistently higher as seen in panel (b) and there is no clear stable minimum in the equation of state. This energy shift can likely be attributed to the suppression of lateral fluctuations of the helium atoms in the presence of corrugation.  For larger values of the chemical potential $\mu > \SI{-80}{\kelvin}$ we begin to see differences between different corrugation levels with a general trend that particle fluctuation effects are enhanced for less rough potentials.  This can be understood in light of the fact that at these higher values of the coverage below $n=1/2$ the adsorbed layer becomes incommensurate, with the strong repulsive interactions between helium atoms becoming dominant and thus reduced corrugation may allow for additional configurational freedom before second layer promotion. 

\begin{figure}[h!]
    \centering
        \includegraphics[width=0.6\linewidth]{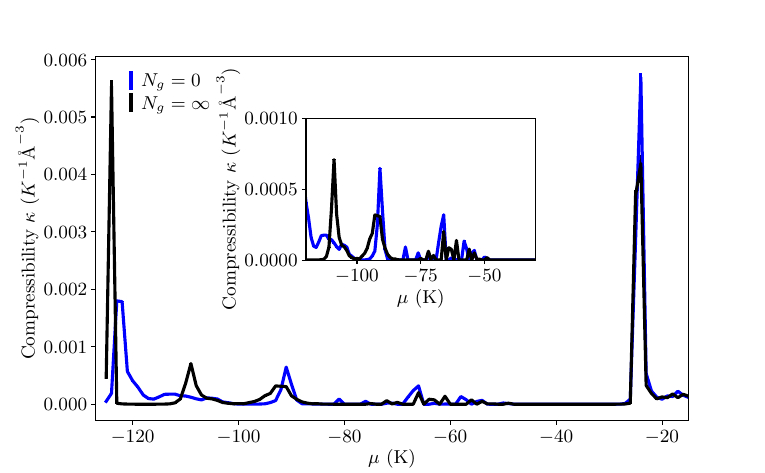}
    \caption{Quantum Monte Carlo results for the compressibility 
 $\kappa$ as a function of chemical potential $\mu$ showing the effects of corrugation measured through the number of reciprocal lattice vectors included in the computation of the adsorption potential $N_g$.  Large spikes accompany layering transitions, or changes between incommensurate and commensurate solids.  The inset shows additional detail of the compressibility for a reduced range of chemical potentials corresponding to the first layer. 
 \label{Fig:5}
 }
\end{figure}

This picture is confirmed in Fig.\ref{Fig:5} which shows the compressibility $\kappa$ for a smooth $N_g=0$ and fully corrugated $N_g=\infty$ potential. 
For the case of corrugation, we see a sharp spike accompanying the onset of the first layer with $n=1/3$ and then the transition to other commensurate fillings with $n=5/12,23/48$ and $7/12$.  The largest signal occurs as the second layer is adsorbed.   Intermediate regions of $\mu$ with $\kappa=0$ highlight the stepwise adsorption process.  For the smooth potential with $N_g=0$, first layer adsorption occurs in an extended region of $\mu$ as seen in the finite compressibility turning on above $\mu = \SI{-120}{\kelvin}$ and extending to $\SI{-100}{\kelvin}$. Interestingly, the transition to $n=23/48$ filling appears sharper for the smooth potential which is confirmed by the step in Fig.~\ref{Fig:3&4}.  For the promotion to the second layer near $\mu \approx \SI{-26}{\kelvin}$, corrugation remarkably still plays a role, with a 50\% larger value of $\kappa$ seen for the smooth potential. 

\subsection{Finite Size Effects}

While the primary focus of this work is to study the effects of corrugation on adsorption in the grand canonical ensemble, the $N_\graphene = 48$ cell analyzed in the previous section is small enough that the stabilization of some particular filling fractions (e.g. $n = 5/12$ and $7/12$) may be construed to be the results of finite size effects as they have not been reported in prior studies of larger cells \cite{bib17, bib23}, where fillings of $7/16$ were prevalent. Therefore, our aim is to provide a more comprehensive depiction of the commensurate and incommensurate phases in the first layer and we conducted  simulations using larger system sizes comprising $N_\graphene = 72$ and $108$ adsorption sites. Where possible, we have extrapolated result (such as the energy per particle) to the thermodynamic limit and attempted to understand the role of a finite cell on the enhanced particle fluctuations observed in Figure \ref{Fig:3&4}.

\begin{figure}[h!]
    \centering
    \includegraphics[width=.8\textwidth]{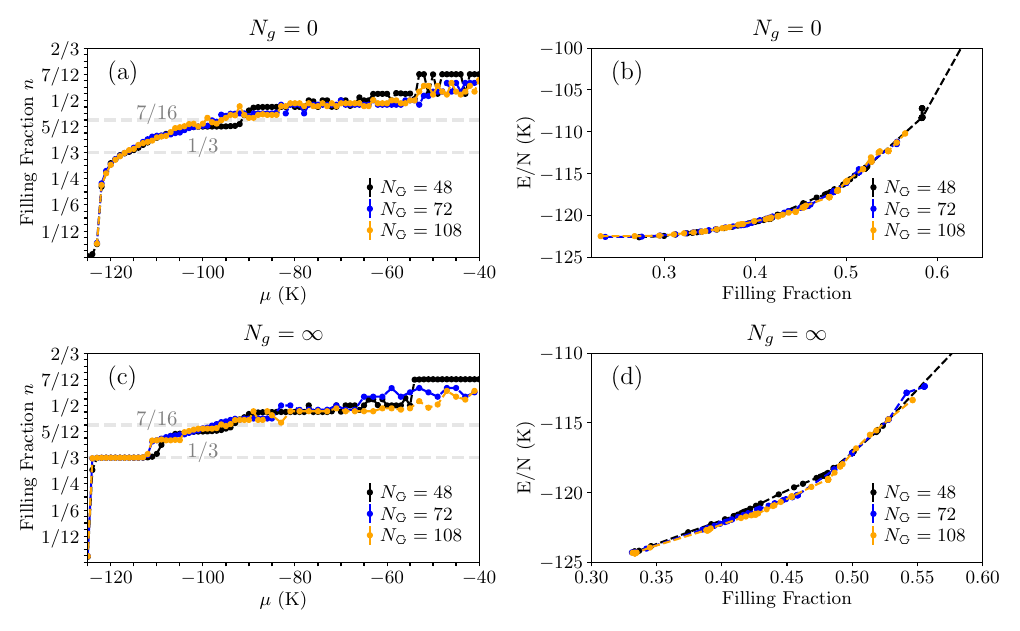}
    \caption{Quantum Monte Carlo results for the filling fraction as a function of chemical potential (a) and the energy per particle (b) for a smooth potential with $N_g=0$ for $N_\graphene = 48, 72,$ and $108$. Panels (b) and (d) show the same quantities for the fully corrugated potential with $N_g = \infty$. Finite size effects are minimal. }
    \label{Fig:9}
\end{figure}
In Fig. \ref{Fig:9}, we compare simulation results at different system sizes focusing on the adsorption process and equation of state for a smooth (panel (a)-(b)) and fully corrugated (panel (c)-(d)) potential.  In the absence of corrugation ($N_g = 0$) we observe minimal finite size effects with the appearance of a smooth onset region of adsorption between $\mu = \SI{-120}{\kelvin}$ and $\mu = \SI{-100}{\kelvin}$. This is supported by the energy per particle in panel (b) which shows no stable minimum across all system sizes and minimal dependence on the value of $N_\graphene$ within statistical fluctuations (error bars are on the order of the symbol size). For $N_g = \infty$, we observe that larger system sizes are accompanied by a steeper first order adsorption transition out of the vacuum to the $\sqrt{3} \times \sqrt{3}$ commensurate solid phase. The filling fraction plateaus at $n=1/3$ in Fig.~\ref{Fig:9} (c) correspond to many points at the minimum value of $E/N \approx 124.34 \pm \SI{0.02}{\kelvin}$ in panel (d) nearly independent of system size within statistical uncertainties. 

\begin{figure}[h!]
    \centering
    \subfloat{\includegraphics[width=0.49\textwidth]{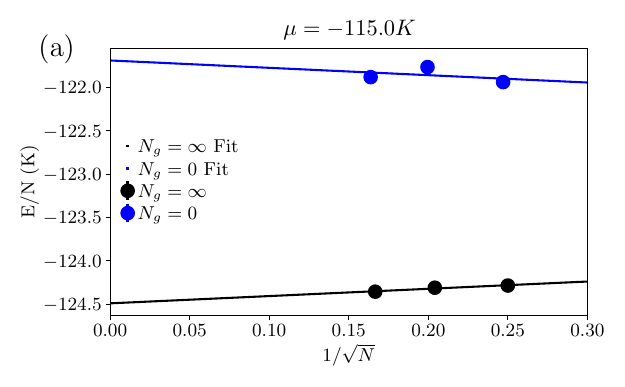}}
    \subfloat{\includegraphics[width=0.49\textwidth]{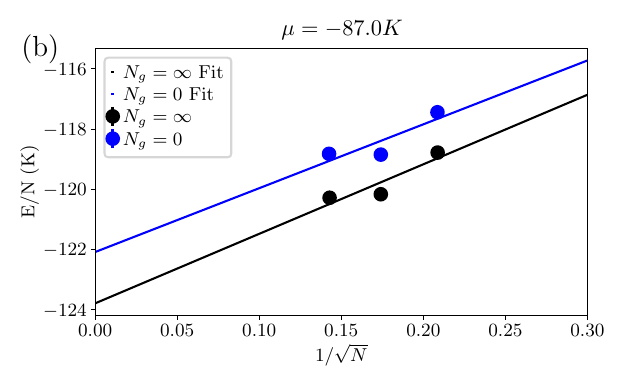}}
    
    \vspace{.01cm}
    \subfloat{\includegraphics[width=0.49\textwidth]{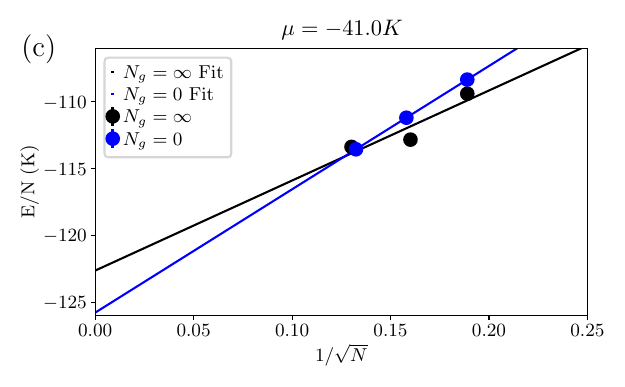}}
    \subfloat{ 
        \vspace{15pt}
        \begin{tabular}{llll}
     \multicolumn{4}{c}{E/N ($N_g = 0$)}\\
    $\mu$ (K) & $N_\graphene = 48$ & $N_\graphene \to \infty$ & \% diff.\\
     \toprule
    $-115$ & -121.941(8) & -121.7(4) & 0.2 \\
    $-87$ & -117.440(9) & -122(2) & 4 \\
    $-41$ & -108.340(6) & -125.79(6) & 14 \\
    \addlinespace[1em]
    \multicolumn{4}{c}{E/N ($N_g = \infty$)}\\
    $\mu$ (K) & $N_\graphene = 48$ & $N_\graphene \to \infty$ & \% diff.\\
    \toprule
    $-115$ & -124.283(5) & -124.49(4) & 0.2\\
    $-87$ & -118.780(9) & -124(2) & 4\\
    $-41$ & -109.397(6) & -123(4) & 11\\
    \end{tabular}
    }
    
    \caption{Finite size scaling.  The energy per particle $E/N$ as a function of $1/\sqrt{N}$ at (a) $\mu = \SI{-115}{\kelvin}$, (b) $\mu = \SI{-87}{\kelvin}$ , and (c) $\mu = \SI{-41}{\kelvin}$(c). The fourth panel shows a table with QMC data for $N_\graphene =48$ and the extrapolated values to the thermodynamic limit. The final number in parenthesis indicates the uncertainty in the final digit.  
}
    \label{Fig:10}
\end{figure}

This is further supported in Fig.~\ref{Fig:10} where we extrapolate the equation of state at fixed chemical potential to the thermodynamic limit using the ansatz:
\begin{equation}
E(N) = E|_\infty N + \mathcal{O}(\sqrt{N})\, .
\end{equation}
At $\mu = \SI{-115}{\kelvin}$, corresponding to the C1/3 phase in the presence of a corrugated potential, the finite size scaling has a minimal negative slope for $N_g = \infty$.  More fluctuations are observed for $N_g = 0$ consistent with the fact that the commensurate phase is not observed in the absence of a corrugated potential (i.e. this is not a special chemical potential). The energy difference between the corrugated and smooth potential is approximately $\SI{2}{\kelvin}$ and is only weakly dependent on system size.  It is also instructive to look at larger values of the chemical potential, where the onset of new incompressible phases becomes apparent. In Fig.~\ref{Fig:9}, we observe a region of stable $n=7/16$ filling around \SI{-100}{\kelvin} for the smooth potential and \SI{-105}{\kelvin} in the presence of corrugation for $N_\graphene = 72$ and $N_\graphene = 108$. At these higher filling fractions, finite-size effects become more prominent. At $\mu = \SI{-87}{\kelvin}$, shown in Fig.~\ref{Fig:10}(b), the change in $E/N$ for $n=7/16$ is approximately 4\% when extrapolating to the thermodynamic limit, with similar scaling behavior observed for both $N_g = 0$ and $N_g = \infty$, with the former always being about \SI{1}{\kelvin} larger. 

Limited finite size effects for the three cells studied are observed in the range $\mu = \SI{-80}{\kelvin}$ to $\mu = \SI{-60}{\kelvin}$ as can be seen in Fig.~\ref{Fig:9} where a filling that fluctuates around $n \approx 1/2$ is observed.  The smallest system size corresponding to $N_\graphene = 48$ does realize some rational values of $n$ not observed for larger sizes due to the appearance of metastable configurations with large finite size energy barriers between them.  As we approach second layer promotion at $\mu = -41$ K, we observe more severe finite-size scaling in Fig.~\ref{Fig:10}(c) where energies per particle in the thermodynamic limit are reduced by nearly 10\%.  This is likely the result of stronger and more important repulsive interactions between helium atoms that are sensitive to small deviations in separation as a result of the emergence of another dominant lengthscale -- the hardcore size of the helium-helium potential.   

\subsection{Second Layer Formation}

\begin{figure}[h!]
    \centering
    \includegraphics[width=.8\textwidth]{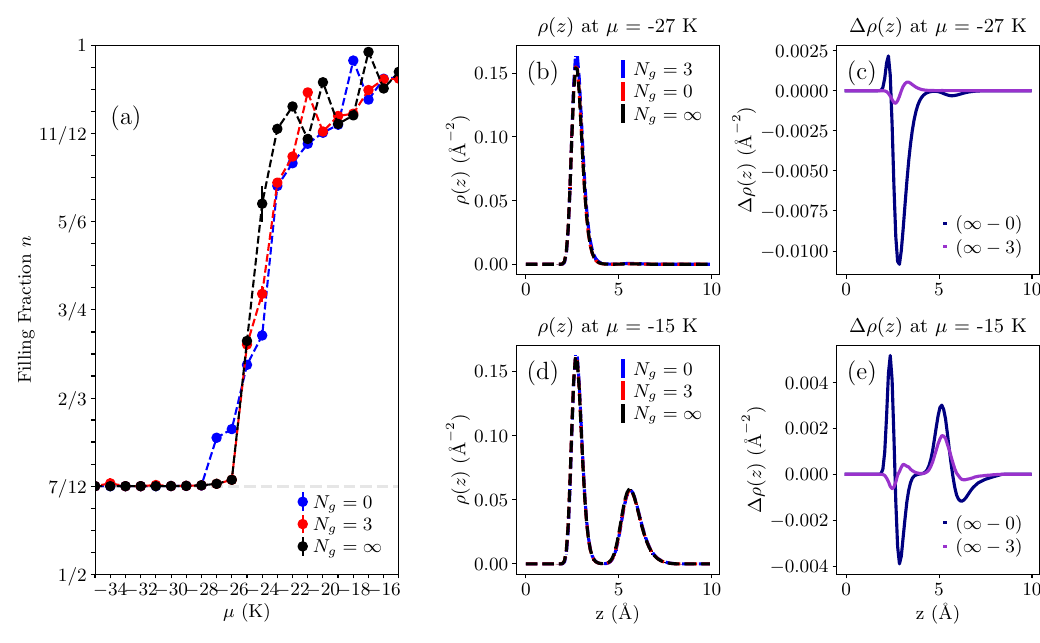}
    \caption{Panel (a) reveals the filling fraction, the number of particles adsorbed over the $N_\graphene = 48$ strong adsorption sites in the graphene membrane. The large increase in the filling fraction is evidence of the adsorption of the second helium layer. Panel (b) and (d) show the linear density at \SI{-15}{\kelvin}  and \SI{-27}{\kelvin}, respectively, with an appearance of a second peak around \SI{6}{\angstrom} for increased chemical potential. Panels (c) and (e) show the difference in the linear density with the full corrugated potential in panels (b) and (d), respectively. 
    }
    \label{Fig:6}
\end{figure}

Details of second layer formation can be observed in Fig. \ref{Fig:6} which shows the chemical potential dependence of the filling fraction for a reduced window of $\mu > -\SI{35}{\kelvin}$. 

Here we see that as the chemical potential is increased, a single layer commensurate solid with filling fraction $n=7/12$ becomes highly compressible as a second layer is formed.  This occurs when there is no longer any energetic advantage from the adsorption potential of adding additional atoms to the first layer, and instead the hard-core helium-helium interaction stabilizes a second adsorbed layer which sits about $z\approx\SI{6.139}{\angstrom}$ from the membrane consistent with previous simulations \cite{bib23}.  The second layer does appear to be starting earlier (near $\mu = \SI{-28}{\kelvin}$) for the smooth potential with $N_g=0$, a feature that was consistent across multiple initial configurations of our simulations. 

Fluctuations in the filling during second layer onset are due to atoms being inserted near $z \approx \SI{6.139}{\angstrom}$ and are not the result of extra atoms stuck above the surface (which could be possible in the grand canonical ensemble). It does appear that for $N_\graphene = 48$, fluctuations are stronger in the presence of corrugation. The averaged transverse structure of the first adsorbed layer $\rho(z)$ directly before second layer formation is shown in panel (b) and looks nearly identical to that seen in Fig.~\ref{Fig:3&4}(a) in the $n=1/3$ commensurate phase, with corrugation effects tending to bind atoms closer and more strongly to the membrane as seen in panel (c) which shows the differences in the averaged transverse density, $\Delta \rho(z)$ for different values of $N_g$.  This is the expected behavior due to the form of the adsorption potential seen in Fig.~\ref{Fig:2} (c). After the second layer is adsorbed, we see multiple features in $\rho(z)$ (panel (d)) with the second layer being much more weakly bound with a larger transverse width.  Remarkably, there are still some structural differences in $\rho(z)$ due to corrugation as quantified in panel (e) with evidence of an enhancement of the density in both layers for $N_g=\infty$. Hence, within this picture, we conjecture that fluctuations in the filling during the second layer onset are a result of the first layer adsorbing new particles, where a stronger corrugated potential can allow for tunneling between different states differing by a few particles as a result of the competition between helium-helium and helium-carbon interactions. 

\begin{figure}[t!]
    \centering
    \includegraphics[width=.45\textwidth]{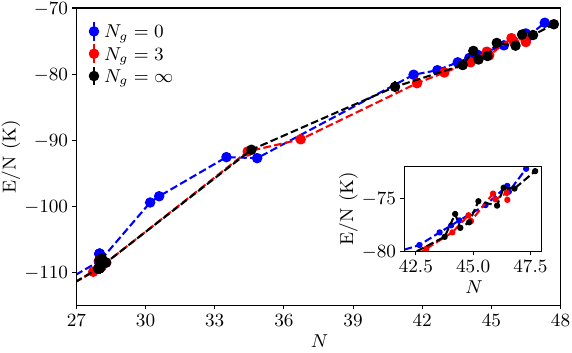}
    \caption{The energy per particle $E/N$ as a function of the total number of particles adsorbed for the same range of chemical potentials seen in Fig.~\ref{Fig:5} corresponding to the second layer onset for different corrugation levels $N_g$ and $N_\graphene = 48$. The inset shows additional detail at the high density. 
    }
    \label{Fig:7}
\end{figure}

Energetics in the grand canonical second layer adsorption process appear complicated as shown in Fig.\ref{Fig:7} and at larger filling fractions there are no clear effects of corrugation that can be discerned. The smooth potential does appear too allow for configurations that are not energetically favorable in the presence of corrugation. 

\section{Discussion}\label{sec13}

In this paper, we presented quantum Monte Carlo results at $T=\SI{1}{\kelvin}$ to study the effects of corrugation of the graphene substrate potential during the helium adsorption process in the grand canonical ensemble.  Our work was motivated by the recent calculations of Boninsegni and Moroni \cite{bib21} who performed analogous simulations in the canonical ensemble for helium on graphite. Here, we considered an adsorption potential consisting of the superposition of isotropic He-C interactions, and systematically included the effects of corrugation near the membrane by modifying the number of reciprocal lattice vector shells included in the substrate potential.  

When no corrugation effects are included, corresponding to $N_g=0$, the adsorption potential is equivalent to that of a uniform film with the same density as graphene and we observe no sharp first layer adsorption transition with a nearly continuous onset of density until a commensurate film with $n=23/48$ is formed above $\mu = \SI{-90}{\kelvin}$. In the presence of corrugation, we instead observe a sharp adsorption transition out of the vacuum to a commensurate $n=1/3$ filled state, where the configuration is stabilized by the large repulsive helium-helium interactions that inhibit any configurations with neighboring triangular lattice sites occupied \cite{Yu:2021gr}.  Other commensurate phases with $n=5/12, 23/48$ and eventually $7/12$ are seen in the first layer, understood to be related to the finite size of the membrane used in the simulations with $N_\graphene = 48$ strong adsorption sites.  When considering larger systems with $N_\graphene = 72$ and $108$, we observe relatively weak finite size effects, with small (order 10\%) modifications in the equation of state when extrapolating to the thermodynamic limit.  These are most severe directly before second layer promotion where the size of the hard core of the helium-helium interaction potential begins to take on a more important energetic role.  The enhanced strength of the corrugated adsorption potential tends to bind the first layer closer to the graphene membrane as compared to the smooth potential, and the energy per particle is lower for all chemical potentials considered in the first layer, regardless of $N_\graphene$. 

Rather surprisingly, as the chemical potential is increased above \SI{-35}{\kelvin} we still observe some effects of corrugation immediately before and during second layer formation. Here, the transverse structure of the first and second layer show modifications of the order of a few percent between $N_g=0$ and $N_g=\infty$.  In addition, for the smooth potential, we find an enhanced compressibility during second layer promotion which may be understood in terms of lowering the energetic barrier to form an incommensurate triangular lattice directly before onset. 

Much remains to be done, including exploring the effects of anisotropy on the He-C potential and understanding the detailed structure and properties of a superfluid film in the second layer in light of experiments on graphite already discussed in the introduction \cite{Nyki:2017py,bib19,bib8}.  The main conclusion of this work is that to achieve quantitative accuracy during the adsorption process of helium on graphene as a function of increased pressure, the knob turned in experiments, including corrugation with at least a few reciprocal lattice vector shells ($N_g\ge 2$) is likely important.  However, to actually achieve realistic agreement with future experiments, more work is needed to understand how accurate simple empirical adsorption potentials actually are, and how advances in quantum chemistry methods may be employed \cite{Moroni:2021lv} to derive ab initio adsorption potentials in a regime dominated by weak van der Waals interactions. 

\backmatter

\section{Acknowledgments}

This research was supported by the National Science Foundation Materials Research Science and Engineering Center program through the UT Knoxville Center for Advanced Materials and Manufacturing (DMR-2309083). The computation for this work was performed on the University of Tennessee Infrastructure for Scientific Applications and Advanced Computing (ISAAC) computational resources.

\section{Availability of Data and Code}
We used an open source path integral quantum Monte Carlo code \cite{bib32} to generate all raw data which is available online \cite{del_maestro_2023_10137837}. The code, analysis and plotting scripts, and processed data needed to reproduce all figures and results in this paper are available in Ref.~\cite{papersCodeRepo}.

\bibliography{bibliography}

\end{document}